\documentclass[10pt]{amsart}

\usepackage[small]{caption}
\usepackage{amsmath,amsfonts,amscd,amssymb,epsf, epsfig}\usepackage{graphicx}

\numberwithin{equation}{section}

\newcommand{\Z}{{\mathbb{Z}}}

\newcommand{\pa}{\partial}

\begin{document}
 
\title{Finite Temperature Casimir Effect in Kaluza-Klein Spacetime}

\author{L.P. Teo}\address{Faculty of Information
Technology, Multimedia University, Jalan Multimedia, Cyberjaya,
63100, Selangor Darul Ehsan, Malaysia.}\email{lpteo@mmu.edu.my}


\begin{abstract}

 In this article, we consider the finite temperature Casimir effect in Kaluza-Klein spacetime due the the vacuum fluctuation of massless scalar field with Dirichlet boundary conditions. We consider the general case where the extra dimensions (internal space) can be any compact connected manifold or orbifold without boundaries. Using piston analysis, we show that the Casimir force is always attractive at any temperature, regardless of the geometry of the internal space. Moreover, the magnitude of the Casimir force increases as the size of the internal space increases and it reduces  to the Casimir force in (3+1)-dimensional Minskowski spacetime when the size of the internal space shrinks to zero. In the other extreme where the internal space is large, the Casimir force can increase beyond all bound.   Asymptotic behaviors of the Casimir force in the low and high temperature regimes are derived and it is observed that the magnitude of the Casimir force grows linearly with temperature in the high temperature regime.

\end{abstract}
\maketitle

\section{ Introduction}
One of the interesting predictions of string theory is that we live in universe with nine or ten space dimensions, three of which are visible. In fact, spacetime with an extra dimension curled up to a tiny circle has already been postulated by Kaluza and Klein \cite{1,2} around 1920's in an attempt to unify two of the fundamental interactions -- gravitational  and electromagnetic forces. Besides the motivation coming from string theory, the interest in universe with more than three spatial dimensions is also stimulated by the developments in particle physics and cosmology. Different spacetime models that contain extra dimensions have been proposed in the endeavor to find a satisfactory explanation for the large hierarchy between some fundamental scales, as well as to account for the dark energy that accelerates the expansion of the universe \cite{3,4,5,6,7,8,9,10,11,12, 28, 32, 65, 67, 33, 30, 66, 29, 31, 13, 14, 39}. One of the proposed form of the dark energy is the cosmological constant -- a constant  energy density   physically equivalent to the vacuum energy or Casimir energy. Postulated  in 1948 \cite{75}, Casimir effect has penetrated into different areas of physics such as quantum field theory, condensed matter physics, atomic and
molecular physics, gravitation and cosmology, and  mathematical physics \cite{76}. In the scenarios of extra dimensional physics, Casimir effect has been  studied in the context of string theory \cite{16, 17, 18, 19}, dark energy and cosmological constant \cite{28, 32, 33, 30, 66, 29, 31, 39, 20, 49, 50, 51, 53, 21, 22, 47},  as well as   stabilization of extra dimensions \cite{ 73, 74, 48, 23, 52, 55, 41, 42, 43, 24, 25, 44, 45, 26, 27, 54}. Recently, Casimir force acting on a pair of parallel plates in Kaluza-Klein spacetime model with $n$ extra dimensions compactified to a $n$-torus $T^n=(S^1)^n$ and in Randall-Sundrum spacetime model have been calculated and analyzed for massless scalar field with Dirichlet boundary conditions on a pair of parallel plates in \cite{34, 35, 36, 68, 37, 38} and in \cite{58, 59, 60} respectively.  For electromagnetic field, the Casimir force   acting on a pair of parallel and  perfectly conducting plates in (4+1)-dimensional Kaluza-Klein spacetime is studied in \cite{39, 46, 56, 57,77}.

Although it was   pointed  out not long after the discovery of Casimir that thermal corrections have to be taken into account in the determination of Casimir effect, majority of the work done for Casimir effect, especially those related to spacetime with extra dimensions, are at zero temperature. This might due to a few reasons.   One of them being the    mathematical techniques required for the computations of Casimir effect at finite temperature is more complicated than those used for zero temperature.  Another might be the fact that the thermal corrections will only be significant at plate separation larger than $1\mu$m. Therefore in the present experimentally accessible measurements of Casimir force \cite{69, 70, 71, 72}, the small separation required for the detectability of Casimir force renders the thermal correction at room temperature negligible. A third reason may be attributed to  the controversial state of the results obtained for finite temperature Casimir force from the point of view of thermodynamics \cite{78}. In any way, by no means    these issues can mask the importance of  taking the thermal correction into account when considering Casimir effect.  In fact, for spacetime with more than four dimensions, the finite temperature Casimir energy   was first calculated in \cite{62} for $(d+1)$-dimensional rectangular cavities   using dimensional regularization. Some other studies that consider finite temperature Casimir effect for higher dimensional spacetime can be found in \cite{48, 55,  1_5_1, 79, 85, 86, 87, 83, 82, 84, 80, 88,  63, 64}. In the context of Kaluza-Klein spacetime model,  the finite temperature Casimir force  has only been investigated in \cite{35} for a pair of parallel plates in spacetime of the form $M^4\times S^1$, where $M^4$ is the (3+1)-- dimensional Minkowski spacetime, and it was claimed that the Casimir force can become repulsive under certain conditions. Recently we re-calculate the Casimir force for the case considered in \cite{35} but using a different regularization setup called the piston approach,  and we find that the  Casimir force should always be attractive \cite{89}.

As a matter of fact, even at zero temperature, except for results for parallel plates, contradictory results for Casimir force   appear due to the employment of different renormalization procedures. The issue on renormalizability of  surface divergence terms   was brought up in \cite{106, 107} and is still under  discussions \cite{44, 109, 90, 110, 91, 92}. In \cite{61}, Cavalcanti proposed a new geometric setup called piston which can avoid the problem. He showed that the Casimir force acting on the piston is free of surface divergence and can therefore be calculated unambiguously. Since then, Casimir piston has attracted considerable interest \cite{37, 93,94, 95,96,97,98,99,100,101,102,103,104,105, 1_15_5}. In fact, the piston scenario has been used in the early work on  Casimir force between parallel plates as a regularization procedure \cite{112,113}.

In this paper, we study the finite temperature Casimir effect in a general Kaluza-Klein spacetime of the  form $M^4\times N^n$, where the internal space $N^n$ can be any $n$-dimensional compact connected manifold or orbifold.   To ensure that the spacetime $M^4\times N^n$ is connected, it is necessary to assume that $N^n$ is connected. Other from this, we do not make any additional assumptions on the topology or geometry of the internal space.  In order to obtain a finite unambiguous result for the Casimir force,   piston setup is used. We   consider massless scalar field with Dirichlet   boundary conditions and derive  exact and explicit formulas for the Casimir force acting on the piston. From this, we draw the conclusion that the Casimir force is always attractive regardless of the temperature. Moreover the Casimir force reduces to the Casimir force in $M^4$ when the size of the internal space goes to zero. This shows that when the internal space becomes negligible, the four dimensional theory is recovered.  Asymptotic behaviors of the Casimir force for different limits such as low and high temperature, small plate separation, small and large extra dimensions are worked out. In the high temperature regime, it is noticed that the magnitude of the Casimir force grows linearly with temperature. This implies that the Casimir force might become significant when the temperature is very high.

Part of the results in this paper has already been  announced in our previous  letter \cite{89}. After this work is done, we notice the closely related work \cite{1_15_6} by Kirsten and Fulling where the zero temperature case is discussed in detail and the finite temperature case appears as a brief remark in the conclusion. In the followings the units with $\hbar=c=k_B=1$ are used throughout.

\section{Basic Formalism}
\begin{figure}\center
\epsfxsize=0.5\linewidth \epsffile{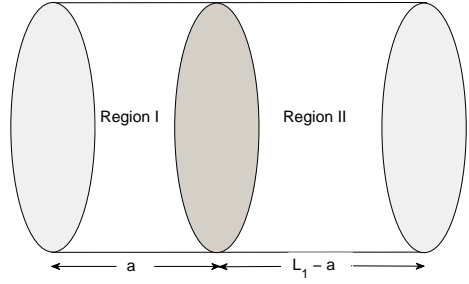} \caption{\label{f1} A movable piston inside a closed cylinder divides the cylinder into two regions.}\end{figure}

We consider Kaluza-Klein spacetime of the form $M^4\times N^n$, where $M^4$ is the (3+1)-dimensional Minkowski spacetime and $N^n$ is the internal space of dimension $n$, assumed to be compact and connected. We are interested in the Casimir effect due to massless scalar field $\varphi$ with Lagrangian  and action
\begin{equation*}
\mathcal{L}=\frac{1}{2} g^{\mu\nu}\partial_{\mu}\varphi \partial_{\nu}\varphi, \hspace{1cm} S=\int \mathcal{L} \sqrt{|g|} d^{d+1}x.
\end{equation*}Here $$ds^2=g_{\mu\nu}dx^{\mu}dx^{\nu}=\eta_{\alpha\beta}dx^{\alpha}dx^{\beta} - G_{ab}dx^adx^b$$ is the spacetime metric with $\eta_{\alpha\beta}=\text{diag}(1, -1, -1, -1)$ the usual (3+1)-$D$ metric on $M^4$;   $ds_N^2=G_{ab}dx^adx^b$ a Riemannian metric on $N^n$; $|g|=(-1)^{n-1}\det[g_{\mu\nu}]$ is the absolute value of the determinant of the matrix $[g_{\mu\nu}]_{\mu,\nu=0}^{n+3}$ and $d=n+3$ is the total space dimension. The field $\varphi$ satisfies the Laplace equation
\begin{equation}\label{eq1_6_1}
\frac{1}{\sqrt{|g|}}\partial_{\mu} \sqrt{|g|}g^{\mu\nu}\partial_{\nu} \varphi=0.
\end{equation}Our aim is to calculate the Casimir force  acting on a pair of parallel plates due to the vacuum fluctuations of the massless scalar field with  Dirichlet   boundary conditions on the plates. As has been pointed out in \cite{38}, a correct regularization approach to this problem is   the piston setup \cite{61, 112}, where the Casimir force was first computed for the case  the two plates are embedded in a closed cylinder \cite{97} and then  the limit where the surrounding cylinder is brought to infinity is evaluated. However since it is enough for us to consider the force acting on one of the plates, it will be sufficient to treat the plate concerned as a movable piston inside a closed cylinder (see FIG. \ref{f1}) and find the Casimir force before letting one end of the cylinder approach infinity. In fact, since in reality we can never have infinite parallel plates, it will be interesting by its own to consider the Casimir force acting on a movable piston inside a closed cylinder. We work with the full generality where the cylinder is allowed to have arbitrary cross section. Mathematically speaking, this means that the cross section of the cylinder is a   simply connected domain $\Omega$ on the plane.
According to \cite{61}, the Casimir energy of the piston system is the sum
\begin{equation*}\begin{split}
E_{\text{Cas}}(a; L_1; T) =& E_{\text{Cas}}^{\text{I}}(a; T) + E_{\text{Cas}}^{\text{II}}(L_1-a; T)  +E_{\text{Cas}}^{\text{ext}}(T) \end{split}
\end{equation*}of the Casimir energies of Region I, Region II and the exterior region, and the Casimir force acting on the piston is given by
\begin{equation}\label{eq1_7_6}
F_{\text{Cas}}(a; L_1; T) =-\frac{\pa}{\pa a}E_{\text{Cas}}(a; L_1; T).
\end{equation} Being independent of the piston position,  the Casimir energy of the exterior region would not contribute to the Casimir force. To compute the Casimir energy in Regions I and II, it suffices to compute the Casimir energy inside the cylinder $I\times \Omega \times N^n$, where $I=[0,L]$, and letting $L$ equal to $a$ and $L_1-a$ respectively. Using mode sum approach, the finite temperature Casimir energy   inside the cylinder $I\times \Omega \times N^n$ is given by
\begin{equation}\label{eq1_6_3}
E_{\text{Cas}}^{\text{cyl }}(L;  T) =-T\log\mathcal{Z}=\frac{1}{2}\sum \omega + T\sum \log\left(1- e^{-  \omega/ T}\right),
\end{equation}where $$\mathcal{Z}=\prod \frac{e^{-\frac{\omega}{2T}}}{1-e^{-\frac{\omega}{T}}}$$ can be regarded as the partition of a grand canonical ensemble, $T$ is the   temperature and the summations in \eqref{eq1_6_3} run through all nonzero eigenfrequencies $\omega$ of the field $\varphi$ satisfying the equation \eqref{eq1_6_1}, with Dirichlet boundary conditions on the boundaries of the cylinder $I\times \Omega \times N^n$.
Using separation of variables, it is immediate to find that   a complete set of solutions to \eqref{eq1_6_1} with Dirichlet boundary conditions on  $I\times \Omega \times N^n$ is given by
\begin{equation}\label{eq1_6_2}\begin{split}
&\varphi_{k, j, l}(\mathbf{x}^{\boldsymbol{\alpha}}, \mathbf{x}^{\mathbf{a}}) =e^{-i\omega_{k,j,l}t}\sin\frac{\pi k x^1}{L} \phi_j(x^2, x^3)\Phi_l(\mathbf{x}^{\mathbf{a}}),\\
&\omega_{k,j,l} =\sqrt{\left(\frac{\pi k}{L}\right)^2+\omega_{\Omega, j}^2+\omega_{N, l}^2}, \hspace{0.5cm}k, j \in \mathbb{N}, l\in \mathbb{N}\cup\{0\},\end{split}
\end{equation}where $\mathbf{x}^{\boldsymbol{\alpha}}=(t, x^1, x^2, x^3)$ and $\mathbf{x}^{\mathbf{a}}=(x^{4}, \ldots, x^{n+3})$ are the coordinates on $M^4$ and $N^n$ respectively; $\phi_j(x^2, x^3), j=1, 2, \ldots$ is a complete set of eigenfunctions of the Laplace operator $\frac{\pa^2}{\pa (x^2)^2}+\frac{\pa^2}{\pa (x^3)^2}$ on $\Omega$ with Dirichlet boundary conditions $\left.\phi_j \right|_{\pa\Omega}=0$ and   eigenvalues $\omega_{\Omega,  j}^2$   arranged so that $0< \omega_{\Omega,   1}^2\leq \omega_{\Omega,   2}^2\leq \ldots$; and $\Phi_l(\mathbf{x}^{\mathbf{a}}), l=0, 1, 2, \ldots$ is a complete set of eigenfunctions of the Laplace operator $\frac{1}{\sqrt{G}}\frac{\pa}{\pa x^a} \sqrt{G} G^{ab}\frac{\pa }{\pa x^b}$ on $N^n$ with eigenvalues $\omega_{N, l}^2$   arranged so that $0=\omega_{N,0}^2<\omega_{N, 1}^2\leq \omega_{N, 2}^2\leq \ldots$. Notice that since $\Omega$ is simply connected, the Laplace operator with Dirichlet boundary conditions does not have zero eigenvalue. On the other hand, since $N^n$ is a compact connected manifold, its Laplace operator has a single zero eigenvalue which corresponds to constant functions.

If instead of Dirichlet boundary conditions, we consider Neumann boundary conditions, we need to  replace the sine function in \eqref{eq1_6_2} by cosine function and let $k$ runs from zero to infinity; and replace $\phi_j(x^2, x^3)$ with $\psi_j(x^2, x^3), j=0, 1,2, \ldots$ which are eigenfunctions of the Laplace operators on $\Omega$ with Neumann boundary conditions $\left.\frac{\pa \psi_j}{\pa \mathbf{n}}\right|_{\pa\Omega}=0$, where $\mathbf{n}$ is the unit vector normal to $\pa\Omega$, and with eigenvalues $\omega_{\Omega,   j}^2$. In this case there is a zero eigenvalue corresponding to constant functions. Therefore for $k=j=l=0$, the term $\omega_{k,j,l}$ is zero and has to be omitted from the summation in \eqref{eq1_6_3}. In the following, we only consider Dirichlet boundary conditions since the case of Neumann boundary conditions can easily be derived analogously.

\section{Casimir force acting on the piston or a pair of parallel plates}\label{sec3}
Since the first summation in the definition of the Casimir energy given by \eqref{eq1_6_3} is divergent, we introduce a  cut-off and compute the small $\lambda$-expansion of
\begin{equation}\label{eq1_6_4}
E_{\text{Cas}}^{\text{cyl }}(\lambda; L;  T) =\frac{1}{2}\sum_{k,j,l} \omega_{k,j,l} e^{-\lambda \omega_{k,j,l}}+ T\sum_{k,j,l} \log\left(1- e^{-  \omega_{k,j,l}/ T}\right),
\end{equation}up to the term $\lambda^0$. The calculations can be done explicitly using zeta functions and heat kernels \cite{1_6_2, 1_6_3, 1_6_4} and we leave it to the appendix. The result is
\begin{equation}\label{eq1_7_5}\begin{split}
E_{\text{Cas}}^{\text{cyl }}(\lambda; L;  T)=&\sum_{i=0}^{n+2} \frac{\Gamma\left(n+4-i\right)}{\Gamma\left(\frac{n+3-i}{2}\right)}\frac{c_{\text{cyl},i}}{\lambda^{n+4-i}}+\frac{ \log[\lambda\mu]-\psi(1)-\log 2+1}{2\sqrt{\pi}}c_{\text{cyl},n+4}\\&-\frac{T}{2}\left(\zeta_{\text{cyl}, T}'(0; L)+\log(\mu^2) \zeta_{\text{cyl}, T}(0; L)\right),\end{split}
\end{equation}where $\mu$ is a normalization constant with dimension length$^{-1}$,\begin{equation}\label{eq1_7_4}
\zeta_{\text{cyl}, T}(s; L) =\sum_{k=1}^{\infty}\sum_{j=1}^{\infty}\sum_{l=0}^{\infty} \sum_{p=-\infty}^{\infty}\left(\omega_{k,j,l}^2+[2\pi p T]^2\right)^{-s}
\end{equation}is a thermal zeta function, and for $i=0, 1,\ldots, n+4$, $c_{\text{cyl}, i}$ are heat kernel coefficients of the Laplace operator on $I\times \Omega\times N$ with Dirichlet boundary conditions. The dependence of $c_{\text{cyl}, i}$ on $L$ is linear (see appendix), i.e. $$c_{\text{cyl}, i} = \frac{L}{2\sqrt{\pi}}c_{\Omega\times N, i} -\frac{1}{2} c_{\Omega\times N, i-1},$$where $c_{\Omega\times N, i}$ are heat kernel coefficients of $\Omega\times N$ and are independent of $L$. Notice that as $\lambda\rightarrow 0^+$, $E_{\text{Cas}}^{\text{cyl }}(\lambda; L; T)$ contains divergence terms of order $\lambda^{-j}$ for $j=1,2,\ldots, n+4$. The leading divergent term $$\frac{\Gamma(n+4)}{\Gamma\left(\frac{n+3}{2}\right)} \frac{c_{\text{cyl}, 0}}{\lambda^{n+4}} =\frac{\Gamma(n+4)}{\Gamma\left(\frac{n+3}{2}\right)}\frac{\text{vol}(I\times\Omega\times N^n)}{(4\pi)^{\frac{n+3}{2}}}\lambda^{-n-4}$$ is the bulk divergence and is usually subtracted away. However, the regularization of the other divergent terms is a highly nontrivial issue. Naive zeta regularization method set all these divergent terms to zero and might lead to  physicality issue. A regularization procedure close in spirit to the piston scenario was introduced in \cite{1_7_1}. However, since our main interest is the Casimir force, we shall not deal   further with this issue. Upon substituting into the definition of Casimir force
\begin{equation*}\begin{split}
F_{\text{Cas}}(a; L_1; T)=&-\frac{\pa}{\pa a}\left( E_{\text{Cas}}^{\text{cyl}}(a; T) + E_{\text{Cas}}^{\text{cyl}}(L_1-a; T)\right)\\=&-\lim_{\lambda\rightarrow 0^+}\frac{\pa}{\pa a}\left( E_{\text{Cas}}^{\text{cyl}}(\lambda;a; T) + E_{\text{Cas}}^{\text{cyl}}(\lambda; L_1-a; T)\right),\end{split}
\end{equation*}we find that since all $\lambda\rightarrow 0^+$ divergent terms are linear in $L$, their contributions to the Casimir force cancel each other and therefore the $\lambda\rightarrow 0^+$ limit is well-defined and is given by
\begin{equation*}
F_{\text{Cas}}(a; L_1; T)=\frac{T}{2}\frac{\pa}{\pa a}\left\{ \zeta_{\text{cyl}, T}'(0; a)+\zeta_{\text{cyl}, T}'(0; L_1-a)\right\}.
\end{equation*}Here we have also used the fact that $\zeta_{\text{cyl}, T}(0; L)=c_{\text{cyl}, n+4} /(2\sqrt{\pi}T)$ is linear in $L$. Using \eqref{eq1_8_1}, we have the explicit formula:
\begin{equation}\label{eq1_7_11_1}
F_{\text{Cas}}(a; L_1; T)=F_{\text{Cas}}^{\infty}(a; T) -F_{\text{Cas}}^{\infty}(L_1-a; T),
\end{equation}where
\begin{equation}\label{eq1_7_10}\begin{split}
F_{\text{Cas}}^{\infty}(a; T)=& \lim_{L_1\rightarrow \infty}F_{\text{Cas}}(a; L_1; T)\\=&\frac{T}{2}\frac{\pa}{\pa a}\sum_{k=1}^{\infty}\sum_{j=1}^{\infty}\sum_{l=0}^{\infty}\sum_{p=-\infty}^{\infty}\frac{1}{k}\exp\left(-2ka \sqrt{\omega_{\Omega, j}^2+\omega_{N,l}^2+[2\pi pT]^2}\right)\\=& -T \sum_{j=1}^{\infty}\sum_{l=0}^{\infty}\sum_{p=-\infty}^{\infty}\frac{\sqrt{\omega_{\Omega, j}^2+\omega_{N,l}^2+[2\pi pT]^2}}{\exp\left(2a \sqrt{\omega_{\Omega, j}^2+\omega_{N,l}^2+[2\pi pT]^2}\right)-1}.
\end{split}\end{equation}Notice that  the expression \eqref{eq1_7_10} is always negative and is an increasing function of $a$. This means that when one end of the closed cylinder is moved to extremely distant place,  the Casimir force  acting on the piston (or two parallel plates inside an infinitely long cylinder) is always attractive and the magnitude of the force decreases as the plate separations increases. Moreover,  the magnitude of the force decreases in exponential rate. For a finite piston inside a closed cylinder, \eqref{eq1_7_11_1} then shows that the Casimir force always tend to pull the piston away from the equilibrium position $x^1=L_1/2$ towards the nearer end and  the magnitude of the Casimir force increases as the piston is farther away from the equilibrium position.

Eq. \eqref{eq1_7_10} can also be regarded as a high temperature expansion of the Casimir force. It shows that in the high temperature regime, the Casimir force is linear in $T$ with leading term
\begin{equation*}
F_{\text{Cas}}^{\infty, T\gg 1}(a; T)\sim -T \sum_{j=1}^{\infty}\sum_{l=0}^{\infty} \frac{\sqrt{\omega_{\Omega, j}^2+\omega_{N,l}^2 }}{\exp\left(2a \sqrt{\omega_{\Omega, j}^2+\omega_{N,l}^2 }\right)-1},
\end{equation*}corresponding to the $p=0$ term. Since this term is independent of the Planck constant $\hbar$, it is sometimes known as the classical limit \cite{1_15_1, 1_15_2, 1_15_3, 1_15_4}.

For the low temperature behavior of the Casimir force, \eqref{eq1_8_9} shows that $F_{\text{Cas}}^{\infty}(a; T)$ can be written as the sum of the zero temperature Casimir force $F_{\text{Cas}}^{\infty}(a; T=0)$ plus the temperature correction $\Delta_T F_{\text{Cas}}^{\infty}(a; T)$, where the zero temperature Casimir force $F_{\text{Cas}}^{\infty}(a; T=0)$ is given by
\begin{equation}\label{eq1_8_11}\begin{split}F_{\text{Cas}}^{\infty}(a; T=0)=&-\frac{1}{2\pi a}\sum_{k=1}^{\infty}\sum_{j=1}^{\infty}\sum_{l=0}^{\infty}  \frac{\sqrt{\omega_{\Omega, j}^2+\omega_{N,l}^2 }}{k} K_{1}\left(2ka \sqrt{\omega_{\Omega, j}^2+\omega_{N,l}^2 }\right)\\&-\frac{1}{\pi}\sum_{k=1}^{\infty}\sum_{j=1}^{\infty}\sum_{l=0}^{\infty}  \left( \omega_{\Omega, j}^2+\omega_{N,l}^2 \right) K_{0}\left(2ka \sqrt{\omega_{\Omega, j}^2+\omega_{N,l}^2 }\right); \end{split}\end{equation}and the temperature correction to the Casimir force $\Delta_T F_{\text{Cas}}^{\infty}(a; T)$ is given by\begin{equation}\label{eq1_9_6}\begin{split}
\Delta_T F_{\text{Cas}}^{\infty}(a; T)=&-\frac{T}{\pi}\sum_{j=1}^{\infty}\sum_{l=0}^{\infty}\sum_{p=1}^{\infty}
\frac{\sqrt{\omega_{\Omega, j}^2+\omega_{N,l}^2}}{p}K_1\left(\frac{p\sqrt{\omega_{\Omega, j}^2+\omega_{N,l}^2}}{T}\right)\\&+\frac{\pi^2}{a^3}\sum_{k=1}^{\infty}\sum_{j=1}^{\infty}\sum_{l=0}^{\infty} \frac{k^2}{\omega_{k,j,l}\left(e^{\frac{\omega_{k,j,l}}{T}}-1\right)}.\end{split}
\end{equation}Although the attractive property of the Casimir force at any temperature has been observed from the compact expression \eqref{eq1_7_10}, it is interesting to remark that the expression for zero temperature Casimir force given by \eqref{eq1_8_11} also manifests the attractive property at zero temperature.

To investigate the behavior of the Casimir force when the cross section $\Omega$ and the internal space $N^n$ is small or large compared to $a$, we define the size variables $r$ and $R$ in the following way:
\begin{equation*}
r:=\sqrt{ \text{Area}(\Omega)}, \hspace{1cm} R:=\left(\text{Vol}(N^n)\right)^{\frac{1}{n}},
\end{equation*} so that $\Omega$ has area $r^2$ and $N^n$ has volume $R^n$. If we rescale the domain $\Omega$  (resp. $N^n$) to $\Omega/r'$ (resp. $N^n/R'$)\footnote{This is equivalent to rescale the metric $G_{ab}dx^adx^b$ to $(r')^{-2}G_{ab}dx^adx^b$.}, then the eigenvalues $\omega_{\Omega, j}^2$ (resp. $\omega_{N,l}^2$) on $\Omega$ (resp. $N^n$) is related to the eigenvalues $\omega_{\Omega/r', j}^2$ (resp. $\omega_{N/R',l}^2$) on $\Omega/r'$  (resp. $N^n/R'$) by $\omega_{\Omega,j}^2= \omega_{\Omega/r',j}^2/(r')^2$ (resp. $\omega_{N,j}^2= \omega_{N/R',j}^2/(R')^2$). Therefore we can define dimensionless variables $\omega_{\Omega, j}'= r\omega_{\Omega, j}$ and $\omega_{N,l}'=R\omega_{N, l}$ so that they are independent of the relative size of $\Omega$ and $N^n$. The expression for Casimir force \eqref{eq1_7_10} can then be rewritten as
\begin{equation}\label{eq1_7_11}
F_{\text{Cas}}^{\infty}(a; T)=   -T \sum_{j=1}^{\infty}\sum_{l=0}^{\infty}\sum_{p=-\infty}^{\infty}\frac{\sqrt{\left(\frac{\omega_{\Omega, j}'}{r}\right)^2+\left(\frac{\omega_{N,l}'}{R}\right)^2+[2\pi pT]^2}}{\exp\left(2a \sqrt{\left(\frac{\omega_{\Omega, j}'}{r}\right)^2+\left(\frac{\omega_{N,l}'}{R}\right)^2+[2\pi pT]^2}\right)-1}.
\end{equation}Since the function
\begin{equation*}
x\mapsto \frac{x}{e^x-1}
\end{equation*}is a decreasing function, it follows immediately from \eqref{eq1_7_11} that as the size of the internal space decreases (i.e. $R$ decreases), the magnitude of the Casimir force decreases. In the limit the internal space vanishes, all the terms with $l\neq 0$ vanishes and the Casimir force reduces to
\begin{equation}\label{eq1_7_12}
F_{\text{Cas}}^{3D,\infty}(a; T)=  -T \sum_{j=1}^{\infty} \sum_{p=-\infty}^{\infty}\frac{\sqrt{\left(\frac{\omega_{\Omega, j}'}{r}\right)^2 +[2\pi pT]^2}}{\exp\left(2a \sqrt{\left(\frac{\omega_{\Omega, j}'}{r}\right)^2 +[2\pi pT]^2}\right)-1},
\end{equation}which is the Casimir force acting on  a pair of parallel plates inside an infinitely long cylinder with cross section $\Omega$ due to massless scalar field with Dirichlet boundary conditions \cite{97}. Taking similar limit to \eqref{eq1_8_11}, one also finds that in the limit of vanishing internal space, the zero temperature Casimir force reduces to the corresponding  zero temperature Casimir force in $(3+1)$-dimensions given by \cite{97}
\begin{equation*}
F_{\text{Cas}}^{3D,\infty}(a; T=0)=-\frac{1}{2\pi a}\sum_{k=1}^{\infty}\sum_{j=1}^{\infty}  \frac{ \omega_{\Omega, j} }{k} K_{1}\left(2ka \omega_{\Omega, j} \right) -\frac{1}{\pi}\sum_{k=1}^{\infty}\sum_{j=1}^{\infty}    \omega_{\Omega, j}^2  K_{0}\left(2ka  \omega_{\Omega, j} \right).
\end{equation*}

Another interesting property we can read from the expression of Casimir force given by \eqref{eq1_7_10} or \eqref{eq1_7_11} is the effect of increasing the number of extra dimensions. The number of extra dimensions can be increased by adding another extra compact space $\mathcal{N}^{n'}$ of dimension $n'$ to $N^n$ so that the internal space becomes $N^n \times \mathcal{N}^{n'}$, a compact manifold of dimension $n+n'$. The spectrum of the Laplace operator on $N^n \times \mathcal{N}^{n'}$ can be written as $\omega_{N, l}^2 +\omega_{\mathcal{N}, l'}^2$, $l, l'=0,1,2,\ldots$, where $\omega_{N, l}^2$ and $\omega_{\mathcal{N}, l'}^2$ are the spectrums of the Laplace operators on $N^n$ and $\mathcal{N}^{n'}$ respectively. Since each term in the summation of  \eqref{eq1_7_10}  is negative, it is then immediate to deduce that adding the dimension of the internal space increase the magnitude of the Casimir force. Moreover, as the size of $\mathcal{N}^{n'}$ shrinks to zero, the Casimir force when the internal space is $N^n \times \mathcal{N}^{n'}$ reduces to the Casimir force   when the internal space is $N^n$. This shows that adding extra dimensions increase the Casimir force hierarchically.

Now we consider the case where the area of the cross section of $\Omega$ is large compared to the plate separation, i.e. the ratio $r/a$ is large. As is derived in the appendix \ref{a2}, when $r/a$ is large, the leading term of the Casimir force is of order $r^2$. Divide the Casimir force by the area $r^2$ of $\Omega$ and taking the limit $r\rightarrow \infty$, we obtain the Casimir force density on a pair of infinite parallel plates. Its high and low temperature expansions   are given respectively by
\begin{equation}\label{eq1_9_3}
\begin{split}
&\mathcal{F}_{\text{Cas}}^{\parallel}(a;T)=\lim_{r\rightarrow \infty}\frac{F_{\text{Cas}}^{\infty}(a;T)}{r^2}\\= & -   \frac{\zeta_R(3)T}{8 \pi a^3}+ \frac{T}{4\pi^{\frac{3}{2}}a^{\frac{3}{2}}} \sum_{k=1}^{\infty} \sum_{\substack{l\in  \mathbb{N}\cup\{0\}, p\in\Z\\ (p, l)\neq (0,0)}}\left(\frac{\sqrt{\omega_{N, l}^2+\left[ 2\pi p T\right]^2}}{k}\right)^{ \frac{3}{2}}K_{ \frac{3}{2}}\left(2ka\sqrt{ \omega_{N, l}^2+\left[ 2\pi p T\right]^2} \right)\\&-\frac{T}{2\pi^{\frac{3}{2}}a^{\frac{1}{2}}}\sum_{k=1}^{\infty} \sum_{\substack{l\in  \mathbb{N}\cup\{0\}, p\in\Z\\ (p, l)\neq (0,0)}}\frac{\left(\sqrt{ \omega_{N, l}^2 +\left[ 2\pi p T\right]^2}\right)^{ \frac{5}{2}}}{k^{\frac{1}{2}}}K_{ \frac{5}{2}}\left(2ka\sqrt{ \omega_{N, l}^2 +\left[ 2\pi p T\right]^2} \right),
\end{split}
\end{equation}and
\begin{equation}\label{eq1_9_5}
\begin{split}
&\mathcal{F}_{\text{Cas}}^{\parallel}(a;T)\\=& -\frac{\pi^2}{480 a^4}-\frac{3}{8\pi^2 a^2}\sum_{k=1}^{\infty}\sum_{l=1}^{\infty}\frac{\omega_{N,l}^2}{k^2}K_2(2ka\omega_{N,l}) -\frac{1}{4\pi^2 a}\sum_{k=1}^{\infty}\sum_{l=1}^{\infty}\frac{\omega_{N,l}^3}{k}K_1(2ka\omega_{N,l})-\frac{\pi^2 T^4}{90}\\&+\frac{\pi T}{2a^3}\sum_{k=1}^{\infty}\sum_{l=0}^{\infty}\sum_{p=1}^{\infty}
\frac{k^2}{p}\exp\left(-\frac{p}{T}\sqrt{\omega_{N,l}^2+\left(\frac{\pi k}{a}\right)^2}\right)-\frac{T^2}{2\pi^2}\sum_{l=1}^{\infty}\sum_{p=1}^{\infty}\frac{\omega_{N,l}^2}{p^2}K_{2}\left( \frac{p\omega_{N,l}}{T}\right).
\end{split}
\end{equation}Neither of these expressions  show manifestly that the Casimir force density is negative. However, since they are obtained as limits of a negative Casimir force,   the Casimir force density acting on a pair of infinite parallel plates in the presence of extra dimensional space is always attractive regardless of the geometry of the internal space.  In the limit the extra dimensions vanish, the Casimir force \eqref{eq1_9_3} and \eqref{eq1_9_5} reduce to
\begin{equation*}
\begin{split}
\mathcal{F}_{\text{Cas}}^{3D,\parallel}(a;T)=&-   \frac{\zeta_R(3)T}{8 \pi a^3}+ \frac{\sqrt{2}T^{\frac{5}{2}}}{ a^{\frac{3}{2}}} \sum_{k=1}^{\infty} \sum_{p=1}^{\infty}\left(\frac{p}{k}\right)^{ \frac{3}{2}}K_{ \frac{3}{2}}\left(4\pi p k Ta\right) \\&-\frac{4\sqrt{2}\pi T^{\frac{7}{2}}}{ a^{\frac{1}{2}}}\sum_{k=1}^{\infty} \sum_{p=1}^{\infty}\frac{p^{ \frac{5}{2}}}{k^{\frac{1}{2}}}K_{ \frac{5}{2}}\left(4\pi p k Ta \right)
\end{split}
\end{equation*}or\begin{equation*}
\begin{split}
\mathcal{F}_{\text{Cas}}^{3D,\parallel}(a;T)= -\frac{\pi^2}{480 a^4} -\frac{\pi^2 T^4}{90}+\frac{\pi T}{2a^3}\sum_{k=1}^{\infty} \sum_{p=1}^{\infty}
\frac{k^2}{p}\exp\left(-\frac{\pi p k}{Ta} \right),
\end{split}
\end{equation*}which are the well known results for the Casimir force density on a pair of infinite parallel plates in (3+1)-dimensional Minskowski spacetime due to massless scalar field with Dirichlet boundary conditions.

As is discussed above, in the limit the internal space vanishes, the Casimir force always reduces to the corresponding Casimir force in $(3+1)$-dimensional spacetime. Therefore, we can write the Casimir force $F_{\text{Cas}}^{\infty}(a; T)$ as the sum of the Casimir force in $(3+1)$-dimensional spacetime $F_{\text{Cas}}^{3D}(a;T)$ plus the correction term $\Delta_{N}F_{\text{Cas}}^{\infty}(a; T)$ due to the presence of the extra dimensional compact space $N^n$. As can be read from \eqref{eq1_7_10}, the correction term is given by
\begin{equation}\label{eq1_9_7}
\begin{split}
\Delta_{N}F_{\text{Cas}}^{\infty}(a; T)=  -T \sum_{j=1}^{\infty}\sum_{l=1}^{\infty}\sum_{p=-\infty}^{\infty}\frac{\sqrt{\omega_{\Omega, j}^2+\omega_{N,l}^2+[2\pi pT]^2}}{\exp\left(2a \sqrt{\omega_{\Omega, j}^2+\omega_{N,l}^2+[2\pi pT]^2}\right)-1}.
\end{split}
\end{equation}Here the $l=0$ term has been omitted. It will be interesting to investigate whether there is a bound for this correction term. For this purpose, it suffices to consider the behavior of this correction term when the size $R$ of the extra dimensional space is large. In fact, this question is also important since the model with large extra dimensions, which is also known as ADD model \cite{8, 10}, has aroused considerable interest as an alternative to explain the weakness of gravity compared to other forces. As before, using the re-scaling $\omega_{N,l} =\omega_{N,l}'/R$, then by the same method as we derive the asymptotic behavior of the Casimir force when $r/a$ is large in appendix \ref{a2}, but with the roles of $\Omega$ and $N^n$ interchanged, we find that when $R/a$ is large, the leading term of $\Delta_{N}F_{\text{Cas}}^{\infty}(a; T)$ is proportional to $R^n$ -- the volume of $N^n$, and is given explicitly by
\begin{equation}\label{eq1_9_7}\begin{split}&\Delta_{N}F_{\text{Cas}}^{\infty}(a; T)\\=&R^n\Biggl\{ \frac{T}{2^n \pi^{\frac{n+1}{2}}a^{\frac{n+1}{2}}} \sum_{k=1}^{\infty} \sum_{j=1}^{\infty}\sum_{p=-\infty}^{\infty}\left(\frac{\sqrt{\omega_{\Omega, j}^2+\left[ 2\pi p T\right]^2}}{k}\right)^{ \frac{n+1}{2}}K_{ \frac{n+1}{2}}\left(2ka\sqrt{ \omega_{\Omega, j}^2+\left[ 2\pi p T\right]^2} \right)\\&-\frac{T}{2^{n-1}\pi^{\frac{n+1}{2}}a^{\frac{n-1}{2}}}\sum_{k=1}^{\infty} \sum_{j=1}^{\infty}\sum_{p=-\infty}^{\infty}\frac{\left(\sqrt{ \omega_{\Omega, j}^2 +\left[ 2\pi p T\right]^2}\right)^{ \frac{n+3}{2}}}{k^{\frac{n-1}{2}}}K_{ \frac{n+3}{2}}\left(2ka\sqrt{ \omega_{\Omega, j}^2 +\left[ 2\pi p T\right]^2} \right)  \Biggr\}\\&+O\left(R^{n-1}\right)\\
=&R^n\Biggl\{ -\frac{n+1}{2^{n+1}\pi^{\frac{n+2}{2}} a^{\frac{n+2}{2}}}\sum_{k=1}^{\infty}\sum_{j=1}^{\infty}\frac{\omega_{\Omega,j}^{\frac{n+2}{2}}}{k^\frac{n+2}{2}}K_{\frac{n+2}{2}}(2ka\omega_{\Omega,j}) -\frac{1}{2^n\pi^{\frac{n+2}{2}} a^{\frac{n}{2}}}\sum_{k=1}^{\infty}\sum_{j=1}^{\infty}\frac{\omega_{\Omega,j}^{\frac{n+4}{2}}}{k^{\frac{n}{2}}}K_{\frac{n}{2}}(2ka\omega_{\Omega,j}) \\&+\frac{ T^{\frac{n-1}{2}}}{2^{\frac{n-1}{2}}\pi^{\frac{n-3}{2}}a^3}\sum_{k=1}^{\infty}\sum_{j=1}^{\infty}\sum_{p=1}^{\infty}
k^2\left(\frac{\sqrt{\omega_{\Omega,j}^2+\left(\frac{\pi k}{a}\right)^2}}{p}\right)^{\frac{n-1}{2}}K_{\frac{n-1}{2}}\left(\frac{p}{T}\sqrt{\omega_{\Omega,j}^2+\left(\frac{\pi k}{a}\right)^2}\right)\\&-\frac{T^{\frac{n+2}{2}}}{2^{\frac{n}{2}}\pi^{\frac{n+2}{2}}} \sum_{j=1}^{\infty}\sum_{p=1}^{\infty}\left(\frac{ \omega_{\Omega,j}}{p}\right)^{\frac{n+2}{2}}K_{\frac{n+2}{2}}\left( \frac{p\omega_{\Omega,j}}{T}\right)\Biggr\}+O\left(R^{n-1}\right).\end{split}\end{equation}This shows that the correction term of the Casimir force due to the presence of extra dimensions can increase beyond all bounds when the size of the extra dimensions is increased. In the zero temperature  limit, the second expression in \eqref{eq1_9_7} shows that the leading correction term of the zero temperature Casimir force when $R/a$ is large is given by\begin{equation*} \begin{split}\Delta_{N}F_{\text{Cas}}^{\infty}(a; T=0)=&R^n\Biggl\{ -\frac{n+1}{2^{n+1}\pi^{\frac{n+2}{2}} a^{\frac{n+2}{2}}}\sum_{k=1}^{\infty}\sum_{j=1}^{\infty}\frac{\omega_{\Omega,j}^{\frac{n+2}{2}}}{k^\frac{n+2}{2}}K_{\frac{n+2}{2}}(2ka\omega_{\Omega,j}) \\&-\frac{1}{2^n\pi^{\frac{n+2}{2}} a^{\frac{n}{2}}}\sum_{k=1}^{\infty}\sum_{j=1}^{\infty}\frac{\omega_{\Omega,j}^{\frac{n+4}{2}}}{k^{\frac{n}{2}}}K_{\frac{n}{2}}(2ka\omega_{\Omega,j})\Biggr\}+O\left(R^{n-1}\right).\end{split}\end{equation*}
For a pair of infinite parallel plates, similar methods (see \eqref{eq1_12_3} and \eqref{eq1_12_4}) show that the leading behavior of the Casimir force density when $R/a$ is large is given by
\begin{equation}\label{eq1_12_5}
\begin{split}
&\mathcal{F}_{\text{Cas}}^{\parallel}(a;T)\\\sim &  R^{n} \Biggl\{-\frac{(n+2)\Gamma\left(\frac{n+3}{2}\right)\zeta_R(n+3)}{(4\pi)^{\frac{n+3}{2}}}\frac{T}{a^{n+3}}+\frac{2^{\frac{1-n}{2}} T^{\frac{n+5}{2}}}{a^{\frac{n+3}{2}}}
\sum_{k=1}^{\infty}\sum_{p=1}^{\infty}\left(\frac{p}{k}\right)^{\frac{n+3}{2}}K_{\frac{n+3}{2}}\left(4\pi kp Ta\right)\\& -\frac{2^\frac{5-n}{2}\pi T^{\frac{n+7}{2}}}{a^{\frac{n+1}{2}}}\sum_{k=1}^{\infty}\sum_{p=1}^{\infty} \frac{p^{\frac{n+5}{2}}}{k^{\frac{n+1}{2}}} K_{\frac{n+5}{2}}\left(4\pi kp Ta\right)\Biggr\}+O(R^{n-1})\end{split}\end{equation}or\begin{equation}\label{eq1_12_5_2}\begin{split}\mathcal{F}_{\text{Cas}}^{\parallel}(a;T) \sim & 
 R^n \Biggl\{-\frac{(n+3)\Gamma\left(\frac{n+4}{2}\right)\zeta_R(n+4)}{(4\pi)^{\frac{n+4}{2}} a^{n+4}}-
\frac{\Gamma\left(\frac{n+4}{2}\right)\zeta_R(n+4)}{\pi^{\frac{n+4}{2}}} T^{n+4}\\& +\frac{\pi T^{\frac{n+1}{2}}}{2^{\frac{n+1}{2}}a^{\frac{n+7}{2}}}\sum_{k=1}^{\infty}\sum_{p=1}^{\infty} \frac{k^{\frac{n+5}{2}}}{p^{\frac{n+1}{2}}}K_{\frac{n+1}{2}}\left(\frac{\pi kp}{Ta}\right) \Biggr\}+O(R^{n-1}).
\end{split}
\end{equation}The expressions in the brackets of \eqref{eq1_12_5} or \eqref{eq1_12_5_2} are actually the Casimir force density acting on a pair of parallel plates in $d=n+3$ dimensional space. Since present experimental results  on Casimir force \cite{69,70,71,72} have verified the effect to be in close agreement with the theoretical result in $(3+1)$-dimensional spacetime, this shows that if present, the extra dimensions must have  size  negligible compared to the plate separations.

\section{Conclusion}
We have investigated the Casimir force in the presence of extra dimensional space which can be any compact connected manifold or orbifold. Exact and explicit formulas for the Casimir force are derived. It is observed that the Casimir force is always attractive regardless of the temperature and the geometry of the  extra dimensions. Moreover, although the Casimir force decreases when the plate separation increases, it increases as the size of the extra dimensions increases, in the rate proportional to the rate the volume of the extra dimensions is increased.   When the temperature is high or the size of the extra dimensions is large, the Casimir force can increase beyond all bounds. Experimental results may be used to set an upper bound to the size of the extra dimensions.

Although we only consider massless scalar field in this article, the qualitative results for electromagnetic  (spin 1) field can be expected to be the same, although the quantitative analysis will be more involved since we have to consider one-forms instead of functions on an arbitrary manifold. In order to employ the Casimir effect as a stabilization mechanism for the extra dimensions, one may need to take into consideration vacuum fluctuations of fermionic fields. This will be discussed elsewhere.

\appendix
\section{Computations of the Casimir energy and Casimir force}\label{a1}
 First we define the   zeta functions
\begin{equation*}\begin{split}
\zeta_{\Omega }(s)=&\sum_{j=1}^{\infty}\omega_{\Omega,  j}^{-2s},\hspace{0.5cm} \zeta_{N}(s) =\sum_{l=1}^{\infty}\omega_{N,l}^{-2s},\\\zeta_{\Omega\times N} =&\sum_{j=1}^{\infty}\sum_{l=0}^{\infty}\left(\omega_{\Omega, j}^2+\omega_{N, l}^2\right)^{-s},\hspace{0.5cm} \zeta_{\text{cyl}}(s) =\sum_{k=1}^{\infty}\sum_{j=1}^{\infty}\sum_{l=0}^{\infty} \omega_{k,j,l}^{-2s}
\end{split}\end{equation*}and the corresponding global heat kernels
\begin{equation*}\begin{split}K_{\Omega}(t)=&\sum_{j=1}^{\infty}e^{-t\omega_{\Omega,  j}^2},\hspace{0.5cm}K_{N}(t)=\sum_{l=1}^{\infty}e^{-t\omega_{N,l}^2},
\\ K_{\Omega\times N}=&\sum_{j=1}^{\infty}\sum_{l=1}^{\infty} e^{-t\left(\omega_{\Omega, j}^2+\omega_{N, l}^2\right)},\hspace{0.5cm}K_{\text{cyl }}(t) =\sum_{k=1}^{\infty}\sum_{j=1}^{\infty}\sum_{l=0}^{\infty} e^{-t \omega_{k,j,l}^2};
\end{split}\end{equation*}which will be used in the computations in this and the following sections. It is well known that (see e.g. \cite{1_6_2, 1_6_3, 1_6_4}) as $t\rightarrow 0^+$, the heat kernel $K(t)$ has the asymptotic expansion
\begin{equation*}
K (t)=\sum_{i=0}^{q-1} c_{i} t^{\frac{i-D}{2}}+O\left( t^{\frac{q-D}{2}}\right),
\end{equation*}where for $\Omega, N,   \Omega\times N, \text{cyl}$, $D=2, n,   n+2, n+3$ respectively. The heat kernel coefficients are geometric invariants of each of the manifolds. In particular, the coefficient $c_0$ of a manifold $\mathcal{M}$ is related to the volume of $\mathcal{M}$ via the relation $$c_{\mathcal{M},0}=\frac{\text{vol}(\mathcal{M})}{(4\pi)^{\frac{D}{2}}}.$$ On the other hand,   the residues of the   zeta function can be expressed in terms of the heat kernel coefficients by
\begin{equation}\label{eq1_8_3}
\text{Res}_{s=\frac{D-i}{2}}\left(\Gamma(s)\zeta(s)\right) = c_{i}, \;\;\;\text{or}\;\;\; \text{Res}_{s=\frac{D-i}{2}}\zeta(s)=\frac{c_{i}}{\Gamma\left(\frac{D-i}{2}\right)}.
\end{equation}For the latter, it means that if $(D-i)/2$ is a non-positive integer, then $\text{Res}_{s=(D-i)/2}\zeta(s)=0$ or equivalently $\zeta(s)$ is regular at $s=(D-i)/2$.
Since
\begin{equation}\label{eq1_7_8}
\sum_{k=1}^{\infty} e^{-t\left(\frac{\pi k}{L}\right)^2} = \frac{L}{2\sqrt{\pi} }\sum_{k=-\infty}^{\infty}t^{-\frac{1}{2}}e^{-\frac{k^2L^2}{t}}-\frac{1}{2} =\frac{L}{2\sqrt{\pi}}t^{-\frac{1}{2}}-\frac{1}{2}+\; \text{exponential decay terms} 
\end{equation}as $t\rightarrow 0^+$, and \begin{equation}\label{eq1_7_9}K_{\text{cyl} }(t)=\sum_{k=1}^{\infty} e^{-t\left(\frac{\pi k}{L}\right)^2} \times K_{\Omega\times N}(t),\end{equation} we find that
\begin{equation}\label{eq1_8_4}
c_{\text{cyl}, i} = \frac{L}{2\sqrt{\pi}}c_{\Omega\times N, i} -\frac{1}{2} c_{\Omega\times N, i-1},
\end{equation}where by convention, $c_{-1}=0$.  Notice that the coefficients $c_{\Omega, i}, c_{N, i}$ and $c_{\Omega\times N, i}$ are independent of $L$, the length of the interval $I$.

Now we compute the small $\lambda$-expansion of the cut-off dependent Casimir energy of the cylinder $I\times\Omega\times N^n$ defined by \eqref{eq1_6_4} up to the term constant in $\lambda$.
 Using standard techniques, we find that up to the term constant in $\lambda$, the zero temperature part is
\begin{equation}\label{eq1_7_1}\begin{split}
&\frac{1}{2}\sum_{k=1}^{\infty}\sum_{j=1}^{\infty} \sum_{l=1}^{\infty} \omega_{k,j,l}e^{-\lambda \omega_{k,j,l}}\\=& -\frac{1}{2}\frac{\pa}{\pa\lambda} \sum_{k=1}^{\infty}\sum_{j=1}^{\infty} \sum_{l=1}^{\infty} e^{-\lambda \omega_{k,j,l}}
=-\frac{1}{2}\frac{\pa}{\pa\lambda} \frac{1}{2\pi i}\int_{\mathrm{c}-i\infty}^{\mathrm{c}+i\infty}\Gamma(z) \lambda^{-z} \zeta_{\text{cyl}}\left(\frac{z}{2}\right)dz\\
=&-\frac{1}{2}\frac{\pa}{\pa\lambda} \left\{\sum_{i=0}^{n+2} \frac{2\Gamma\left(n+3-i\right)}{\Gamma\left(\frac{n+3-i}{2}\right)}\frac{c_{\text{cyl},i}}{\lambda^{n+3-i}}+c_{n+3}+\lambda\left(\frac{\psi(2)-\log\lambda}{\sqrt{\pi}}c_{\text{cyl},n+4}
-\text{FP}_{s=\frac{-1}{2}}\zeta_{\text{cyl}}(s)\right)\right\}\\
=&\sum_{i=0}^{n+2} \frac{\Gamma\left(n+4-i\right)}{\Gamma\left(\frac{n+3-i}{2}\right)}\frac{c_{\text{cyl},i}}{\lambda^{n+4-i}}-\frac{\psi(1)-\log\lambda}{2\sqrt{\pi}}c_{\text{cyl},n+4}
+\frac{1}{2}\text{FP}_{s=\frac{-1}{2}}\zeta_{\text{cyl}}(s).
\end{split}\end{equation}Here $$\text{FP}_{s=s_0}f(s) = \lim_{s\rightarrow s_0}\left(f(s)-\frac{\text{Res}_{s=s_0} f(s)}{s-s_0}\right)$$ is the finite part of the meromorphic function $f(s)$ at $s=s_0$.
To take into account the thermal correction of the Casimir energy, we consider the thermal zeta functions which are defined by adding   an extra dimension corresponding to a circle of length $1/T$ to each of the spaces considered above. For the cylinder $I\times\Omega\times N^n$, the thermal zeta function is defined by \eqref{eq1_7_4}.
It is standard to find that
\begin{equation}\label{eq1_8_5}\begin{split}
&\zeta_{\text{cyl}, T}(s)=\zeta_{\text{cyl}, T}(s; L) =\frac{1}{\Gamma(s)}\int_0^{\infty} t^{s-1} K_{\text{cyl}}(t) \sum_{p=-\infty}^{\infty} e^{-t[2\pi p T]^2}dt \\=& \frac{1}{2\sqrt{\pi}T}\frac{1}{\Gamma(s)}\int_0^{\infty} t^{s-\frac{1}{2}-1}K_{\text{cyl}}(t) \sum_{p=-\infty}^{\infty}e^{-\frac{1}{t}\left[\frac{p}{2T}\right]^2} dt\\
=&\frac{1}{2\sqrt{\pi}T}\frac{\Gamma\left(s-\frac{1}{2}\right)}{\Gamma(s)}\zeta_{\text{cyl}}\left(s-\frac{1}{2}\right)+\frac{2}{\sqrt{\pi}T}
\frac{1}{\Gamma(s)}\sum_{k=1}^{\infty}\sum_{j=1}^{\infty}\sum_{l=0}^{\infty}\sum_{p=1}^{\infty}\left(\frac{p}{2T \omega_{k,j,l}}\right)^{s-\frac{1}{2}}K_{s-\frac{1}{2}}\left(\frac{p \omega_{k,j,l}}{T}\right),
\end{split}\end{equation}which gives  $\zeta_{\text{cyl}, T}(0)=c_{\text{cyl}, n+4}/(2\sqrt{\pi}T)$ and\begin{equation}\label{eq1_7_2}
\zeta_{\text{cyl}, T}'(0)= -\frac{1}{T}\left( \frac{\log 2-1}{\sqrt{\pi}}c_{\text{cyl}, n+4}+\text{FP}_{s=\frac{-1}{2}}\zeta_{\text{cyl}}(s) \right)-2\sum_{k=1}^{\infty}\sum_{j=1}^{\infty}\sum_{l=0}^{\infty}\log\left(1-e^{-\frac{\omega_{k,j,l}}{T}}\right).
\end{equation}Together with \eqref{eq1_7_1}, we obtain
\begin{equation}\label{eq1_9_1}\begin{split}
&E_{\text{Cas}}^{\text{cyl}}(\lambda; L;  T) =\frac{1}{2}\sum_{k,j,l} \omega_{k,j,l} e^{-\lambda \omega_{k,j,l}}+ T\sum_{k,j,l} \log\left(1- e^{-  \omega_{k,j,l}/ T}\right)\\
=&\sum_{i=0}^{n+2} \frac{\Gamma\left(n+4-i\right)}{\Gamma\left(\frac{n+3-i}{2}\right)}\frac{c_{\text{cyl},i}}{\lambda^{n+4-i}}+\frac{ \log[\lambda\mu]-\psi(1)-\log 2+1}{2\sqrt{\pi}}c_{\text{cyl},n+4}\\&-\frac{T}{2}\left(\zeta_{\text{cyl}, T}'(0)+\log(\mu^2) \zeta_{\text{cyl}, T}(0)\right),
\end{split}\end{equation}where  terms of higher orders in $\lambda$ has been omitted. Here $\mu$ is a normalization constant with dimension length$^{-1}$.

As discussed in section \ref{sec3},  we can use \eqref{eq1_9_1} to show that the Casimir force acting on the piston in the limit $L_1\rightarrow \infty$ is   given by
\begin{equation}\label{eq1_8_6}
F_{\text{Cas}}^{\infty}(a;   T)=\frac{T}{2}\lim_{L_1\rightarrow \infty}\frac{\pa}{\pa a}\left( \zeta_{\text{cyl}, T}'(0; a)+\zeta_{\text{cyl}, T}'(0; L_1-a)\right),
\end{equation}
where    a formula for the function $\zeta_{\text{cyl}, T}'(0;L)$ is given by \eqref{eq1_7_2}. The term $\text{FP}_{s=\frac{-1}{2}}\zeta_{\text{cyl}}(s)  $ in \eqref{eq1_7_2} can be computed more explicitly in the following way. Using \eqref{eq1_7_8} and \eqref{eq1_7_9}, we have
\begin{equation}\label{eq1_8_2}\begin{split}
\zeta_{\text{cyl}}(s)=&\frac{1}{\Gamma(s)}\int_0^{\infty}t^{s-1}K_{\text{cyl}}(t)dt\\=&\frac{1}{\Gamma(s)}\int_0^{\infty}t^{s-1}K_{\Omega\times N}(t)\left(  \frac{L}{2\sqrt{\pi} }\sum_{k=-\infty}^{\infty}t^{-\frac{1}{2}}e^{-\frac{k^2L^2}{t}}-\frac{1}{2}\right)dt\\
=&-\frac{1}{2}\zeta_{\Omega\times N}(s) +\frac{L}{2\sqrt{\pi}}\frac{\Gamma\left(s-\frac{1}{2}\right)}{\Gamma(s)} \zeta_{\Omega\times N}\left(s-\frac{1}{2}\right)\\&+\frac{2L}{\sqrt{\pi}\Gamma(s)}\sum_{k=1}^{\infty}\sum_{j=1}^{\infty}\sum_{l=0}^{\infty} \left(\frac{kL}{\sqrt{\omega_{\Omega, j}^2+\omega_{N,l}^2 }}\right)^{s-\frac{1}{2}}K_{s-\frac{1}{2}}\left(2kL \sqrt{\omega_{\Omega, j}^2+\omega_{N,l}^2 }\right).
\end{split}
\end{equation}This gives
\begin{equation}\label{eq1_8_7}
\begin{split}
&\text{FP}_{s=\frac{-1}{2}}\zeta_{\text{cyl}}(s)\\=&-\frac{1}{2}\text{FP}_{s=\frac{-1}{2}}\zeta_{\Omega\times N}(s)+\frac{L}{4\pi}\left( \psi\left(-\frac{1}{2}\right)c_{\Omega\times N, n+4} -\text{FP}_{s=-1}\left\{\Gamma(s)\zeta_{\Omega\times N}(s)\right\}\right)\\&-\frac{1}{\pi}\sum_{k=1}^{\infty}\sum_{j=1}^{\infty}\sum_{l=0}^{\infty}  \frac{\sqrt{\omega_{\Omega, j}^2+\omega_{N,l}^2 }}{k} K_{1}\left(2kL \sqrt{\omega_{\Omega, j}^2+\omega_{N,l}^2 }\right).
\end{split}
\end{equation}

Besides the formula \eqref{eq1_7_2}, there is an alternative explicit formula for $\zeta_{\text{cyl}, T}'(0;L)$. As in \eqref{eq1_8_2}, we have
\begin{equation*}
\begin{split}
\zeta_{\text{cyl}, T}(s)=&  -\frac{1}{2} \zeta_{\Omega\times N, T}(s) +\frac{L}{2\sqrt{\pi}}\frac{\Gamma\left(s-\frac{1}{2}\right)}{\Gamma\left(s\right)} \zeta_{\Omega\times N, T}\left(s-\frac{1}{2}\right)\\&+\frac{2L}{\sqrt{\pi}\Gamma(s)}\sum_{k=1}^{\infty}\sum_{j=1}^{\infty}\sum_{l=0}^{\infty}\sum_{p=-\infty}^{\infty} \left(\frac{kL}{\sqrt{\omega_{\Omega, j}^2+\omega_{N,l}^2+[2\pi pT]^2}}\right)^{s-\frac{1}{2}}\\&\times K_{s-\frac{1}{2}}\left(2kL \sqrt{\omega_{\Omega, j}^2+\omega_{N,l}^2+[2\pi pT]^2}\right).
\end{split}
\end{equation*}
This gives
\begin{equation}\label{eq1_8_1}
\begin{split}\zeta_{\text{cyl}, T}'(0)=\Lambda_0+\Lambda_1 L + \sum_{k=1}^{\infty}\sum_{j=1}^{\infty}\sum_{l=0}^{\infty}\sum_{p=-\infty}^{\infty}\frac{1}{k}\exp\left(-2kL \sqrt{\omega_{\Omega, j}^2+\omega_{N,l}^2+[2\pi pT]^2}\right),
\end{split}
\end{equation}where $\Lambda_0$ and $\Lambda_1$ are independent of $L$, and $\Lambda_1$ is given by
\begin{equation*}\begin{split}
\Lambda_1 =& -\frac{\psi(1)}{2\sqrt{\pi}}\text{Res}_{s=-\frac{1}{2}}\left\{\Gamma\left(s\right)\zeta_{\Omega\times N, T}(s)\right\} +\frac{1}{2\sqrt{\pi}}\text{FP}_{s=-\frac{1}{2}}\left\{\Gamma\left(s\right)\zeta_{\Omega\times N, T}(s)\right\}. \end{split}\end{equation*}Using \eqref{eq1_8_3}, one can show that
\begin{equation*}
\text{Res}_{s=-\frac{1}{2}}\left\{\Gamma\left(s\right)\zeta_{\Omega\times N, T}(s)\right\} =\frac{1}{2\sqrt{\pi} T}c_{\Omega\times N, n+4}.
\end{equation*}On the other hand, similar to \eqref{eq1_8_5}, we can show that
\begin{equation*}
\begin{split}&\zeta_{\Omega\times N, T} (s)  =\frac{1}{2\sqrt{\pi}T}\frac{\Gamma\left(s-\frac{1}{2}\right)}{\Gamma(s)}\zeta_{\Omega\times N}\left(s-\frac{1}{2}\right)\\&+\frac{2}{\sqrt{\pi}T}
\frac{1}{\Gamma(s)} \sum_{j=1}^{\infty}\sum_{l=0}^{\infty}\sum_{p=1}^{\infty}\left(\frac{p}{2T\sqrt{\omega_{\Omega, j}^2+\omega_{N,k}^2}}\right)^{s-\frac{1}{2}}K_{s-\frac{1}{2}}\left(\frac{p \sqrt{\omega_{\Omega, j}^2+\omega_{N,k}^2}}{T}\right).
\end{split}
\end{equation*}Therefore,
\begin{equation*}
\begin{split}
\text{FP}_{s=-\frac{1}{2}}\left\{\Gamma\left(s\right)\zeta_{\Omega\times N, T}(s)\right\}=&\frac{1}{2\sqrt{\pi}T} \text{FP}_{s=-1}\left\{\Gamma\left(s\right)\zeta_{\Omega\times N}(s)\right\}\\&+\frac{4}{\sqrt{\pi}}\sum_{j=1}^{\infty}\sum_{l=0}^{\infty}\sum_{p=1}^{\infty}
\frac{\sqrt{\omega_{\Omega, j}^2+\omega_{N,l}^2}}{p}K_1\left(\frac{p\sqrt{\omega_{\Omega, j}^2+\omega_{N,l}^2}}{T}\right)
\end{split}
\end{equation*}
and
\begin{equation}\label{eq1_8_8}\begin{split}
\Lambda_1= &-\frac{\psi(1)}{4\pi T}c_{\Omega\times N, n+4}+\frac{1}{4\pi T}\text{FP}_{s=-1}\left\{\Gamma(s)\zeta_{\Omega\times N}(s)\right\}\\&+\frac{2}{\pi}\sum_{j=1}^{\infty}\sum_{l=0}^{\infty}\sum_{p=1}^{\infty}
\frac{\sqrt{\omega_{\Omega, j}^2+\omega_{N,l}^2}}{p}K_1\left(\frac{p\sqrt{\omega_{\Omega, j}^2+\omega_{N,l}^2}}{T}\right).
\end{split}\end{equation}

Using \eqref{eq1_7_2} and \eqref{eq1_8_1} to compute the first  and second terms in \eqref{eq1_8_6}, together with the help of the formulas \eqref{eq1_8_4}, \eqref{eq1_8_7} and \eqref{eq1_8_8}, we find after some simplifications that in the limit $L_1\rightarrow \infty$, the Casimir force is given by
\begin{equation}\label{eq1_8_9}
\begin{split}
F_{\text{Cas}}^{\infty}(a; T) =& -\frac{1}{2\pi a}\sum_{k=1}^{\infty}\sum_{j=1}^{\infty}\sum_{l=0}^{\infty}  \frac{\sqrt{\omega_{\Omega, j}^2+\omega_{N,l}^2 }}{k} K_{1}\left(2ka \sqrt{\omega_{\Omega, j}^2+\omega_{N,l}^2 }\right)\\&-\frac{1}{\pi}\sum_{k=1}^{\infty}\sum_{j=1}^{\infty}\sum_{l=0}^{\infty}  \left( \omega_{\Omega, j}^2+\omega_{N,l}^2 \right) K_{0}\left(2ka \sqrt{\omega_{\Omega, j}^2+\omega_{N,l}^2 }\right)\\
&-\frac{T}{\pi}\sum_{j=1}^{\infty}\sum_{l=0}^{\infty}\sum_{p=1}^{\infty}
\frac{\sqrt{\omega_{\Omega, j}^2+\omega_{N,l}^2}}{p}K_1\left(\frac{p\sqrt{\omega_{\Omega, j}^2+\omega_{N,l}^2}}{T}\right)\\&+\frac{\pi^2}{a^3}\sum_{k=1}^{\infty}\sum_{j=1}^{\infty}\sum_{l=0}^{\infty} \frac{k^2}{\omega_{k,j,l}\left(e^{\frac{\omega_{k,j,l}}{T}}-1\right)}.
\end{split}
\end{equation}

\section{Asymptotic behaviors of the Casimir force}\label{a2}

In this section, we first derive two formulas for the asymptotic behavior of the Casimir force $F_{\text{Cas}}^{\infty}(a; T)$ when the ratio $r/a$ is large. We then derive the asymptotic behavior of the Casimir force  $F_{\text{Cas}}^{\infty}(a; T)$ when both $r/a$ and $R/a$ are large.

First, using the formula \eqref{eq1_7_10} for $F_{\text{Cas}}^{\infty}(a; T)$ and the formula
\begin{equation*}\begin{split}
&\sum_{k=1}^{\infty}  \frac{1}{k}\exp\left(-2ka z\right)=\frac{a}{\sqrt{\pi}} \int_0^{\infty} t^{ \frac{1}{2}-1} \sum_{k=1}^{\infty} \exp\left\{-\frac{z^2}{t} -t k^2a^2 \right\}dt,\end{split}
\end{equation*}we find that
\begin{equation}\label{eq1_9_2}
\begin{split}
&F_{\text{Cas}}^{\infty}(a; T)=\frac{T}{2\sqrt{\pi}}\frac{\pa}{\pa a} \Biggl\{a \int_0^{\infty} t^{ \frac{1}{2}-1} \sum_{k=1}^{\infty}\sum_{j=1}^{\infty}\sum_{l=0}^{\infty}\sum_{p=-\infty}^{\infty}\\& \hspace{2cm}\times\exp\left\{-\frac{1}{t}\left(\left[\frac{\omega_{\Omega, j}' }{r}\right]^2 + \omega_{N, l}^2+\left[2\pi p T\right]^2\right)-t k^2a^2 \right\}dt\Biggr\}\\
=&\frac{T}{2\sqrt{\pi}}\frac{\pa}{\pa a} \Biggl\{a \int_0^{\infty} t^{ \frac{1}{2}-1} \sum_{k=1}^{\infty}\sum_{j=1}^{\infty}\sum_{l=0}^{\infty}\sum_{p=-\infty}^{\infty}\frac{1}{2\pi i}\int_{\mathrm{c}-i\infty}^{\mathrm{c}+i\infty}\Gamma(z) t^z r^{2z}(\omega_{\Omega, j}')^{-2z}\\&\hspace{2cm}\times \exp\left\{-\frac{1}{t}\left(  \omega_{N, l}^2 +\left[ 2\pi p T\right]^2\right)-t k^2a^2 \right\}dz dt\Biggr\}
\\=&\frac{T}{2\sqrt{\pi}}\frac{\pa}{\pa a} \Biggl\{  \frac{a}{2\pi i}\int_{\mathrm{c}-i\infty}^{\mathrm{c}+i\infty}\Gamma(z)   r^{2z}\zeta_{\Omega/r}(z)\Biggl[ \frac{\Gamma\left(z+\frac{1}{2}\right)}{a^{2z+1}}\zeta_R(2z+1)\\& + 2 \sum_{k=1}^{\infty} \sum_{\substack{l\in  \mathbb{N}\cup\{0\}, p\in\Z\\ (p, l)\neq (0,0)}}\left(\frac{\sqrt{ \omega_{N, l}^2 +\left[ 2\pi p T\right]^2}}{ka}\right)^{z+\frac{1}{2}}K_{z+\frac{1}{2}}\left(2ka\sqrt{  \omega_{N, l}^2 +\left[ 2\pi p T\right]^2} \right)\Biggr]dz \Biggr\}\\
=&\frac{T}{2\sqrt{\pi}}\frac{\pa}{\pa a} \Biggl\{a r^2c_{\Omega/r, 0}\Biggl[ \frac{\sqrt{\pi}}{2}\frac{\zeta_R(3)}{a^3}\\&+ 2 \sum_{k=1}^{\infty} \sum_{\substack{l\in \mathbb{N}\cup\{0\}, p\in\Z\\ (p, l)\neq (0,0)}}\left(\frac{\sqrt{ \omega_{N, l}^2 +\left[ 2\pi p T\right]^2}}{ka}\right)^{ \frac{3}{2}}K_{ \frac{3}{2}}\left(2ka\sqrt{ \omega_{N, l}^2 +\left[ 2\pi p T\right]^2} \right) \Biggr]\Biggr\}+O(r/a)\\
=&r^2\Biggl\{ -   \frac{\zeta_R(3)T}{8 \pi a^3}+ \frac{T}{4\pi^{\frac{3}{2}}a^{\frac{3}{2}}} \sum_{k=1}^{\infty} \sum_{\substack{l\in \mathbb{N}\cup\{0\}, p\in\Z\\ (p, l)\neq (0,0)}}\left(\frac{\sqrt{\omega_{N, l}^2+\left[ 2\pi p T\right]^2}}{k}\right)^{ \frac{3}{2}}K_{ \frac{3}{2}}\left(2ka\sqrt{ \omega_{N, l}^2+\left[ 2\pi p T\right]^2} \right)\\&-\frac{T}{2\pi^{\frac{3}{2}}a^{\frac{1}{2}}}\sum_{k=1}^{\infty} \sum_{\substack{l\in \widehat{\mathbb{N}}, p\in\Z\\ (p, l)\neq (0,0)}}\frac{\left(\sqrt{ \omega_{N, l}^2 +\left[ 2\pi p T\right]^2}\right)^{ \frac{5}{2}}}{k^{\frac{1}{2}}}K_{ \frac{5}{2}}\left(2ka\sqrt{ \omega_{N, l}^2 +\left[ 2\pi p T\right]^2} \right)  \Biggr\}+O(r/a).
\end{split}
\end{equation}

For the second formula, \eqref{eq1_8_9} gives
\begin{equation}\label{eq1_12_2}\begin{split}
&F_{\text{Cas}}^{\infty}(a; T) =\frac{\pa }{\pa a}\Biggl\{\frac{a}{4\pi}\int_0^{\infty} \sum_{k=1}^{\infty}\sum_{j=1}^{\infty}\sum_{l=0}^{\infty}\exp\left(-\frac{1}{t}\left(\left[\frac{\omega_{\Omega,j}'}{r}\right]^2+\omega_{N, l}^2\right)-tk^2a^2\right)dt\\&+\frac{1}{2\sqrt{\pi}}\int_0^{\infty} t^{\frac{1}{2}-1}\sum_{k=1}^{\infty}\sum_{j=1}^{\infty}\sum_{l=0}^{\infty} \sum_{p=1}^{\infty} \exp\left(-\frac{1}{t}\left(\left[\frac{\pi k }{a}\right]^2+\left[\frac{\omega_{\Omega,j}'}{r}\right]^2+\omega_{N,l}^2\right)-\frac{tp^2}{4T^2}\right)dt\Biggr\}\\
&-\frac{1}{4\pi}\int_0^{\infty}  \sum_{j=1}^{\infty}\sum_{l=0}^{\infty}\sum_{p=1}^{\infty} \exp\left(-\frac{1}{t}\left( \left[\frac{\omega_{\Omega,j}'}{r}\right]^2+\omega_{N,l}^2\right)-\frac{tp^2}{4T^2}\right)dt.
\end{split}\end{equation}Using the same method as we derive \eqref{eq1_9_2}, we find that
\begin{equation}\label{eq1_9_4}
\begin{split}
F_{\text{Cas}}^{\infty}(a; T) =&r^2\Biggl\{-\frac{\pi^2}{480 a^4}-\frac{3}{8\pi^2 a^2}\sum_{k=1}^{\infty}\sum_{l=1}^{\infty}\frac{\omega_{N,l}^2}{k^2}K_2(2ka\omega_{N,l}) -\frac{1}{4\pi^2 a}\sum_{k=1}^{\infty}\sum_{l=1}^{\infty}\frac{\omega_{N,l}^3}{k}\\&\times K_1(2ka\omega_{N,l}) -\frac{\pi^2 T^4}{90}+\frac{\pi T}{2a^3}\sum_{k=1}^{\infty}\sum_{l=0}^{\infty}\sum_{p=1}^{\infty}
\frac{k^2}{p}\exp\left(-\frac{p}{T}\sqrt{\omega_{N,l}^2+\left(\frac{\pi k}{a}\right)^2}\right)\\&-\frac{T^2}{2\pi^2}\sum_{l=1}^{\infty}\sum_{p=1}^{\infty}\frac{\omega_{N,l}^2}{p^2}K_{2}\left( \frac{p\omega_{N,l}}{T}\right)\Biggr\}+O(r/a).
\end{split}
\end{equation}

\eqref{eq1_9_2} and \eqref{eq1_9_4} show that when the ratio $r/a$ is large, the leading term of the Casimir force is of order $r^2$. To investigate the leading behavior of this term when the ratio $R/a$ is large, we infer from the second, third and fourth equality of \eqref{eq1_9_2} that
\begin{equation} \label{eq1_12_1}
\begin{split}
F_{\text{Cas}}^{\infty}(a; T) =& r^2 \frac{\pa}{\pa a} \left\{\frac{Ta}{8\pi^{\frac{3}{2}}} \int_0^{\infty} t^{ \frac{1}{2}} \sum_{k=1}^{\infty} \sum_{l=0}^{\infty}\sum_{p=-\infty}^{\infty} \exp\left\{-\frac{1}{t}\left(  \omega_{N, l}^2 +\left[ 2\pi p T\right]^2\right)-t k^2a^2 \right\}  dt\right\}\\&+O(r/a).
\end{split}
\end{equation}Using the same method as we derive \eqref{eq1_9_2}, we find from \eqref{eq1_12_1}  that when $R/a$ is large, the term of order $r^2$ of the Casimir force $F_{\text{Cas}}^{\infty}(a; T)$ given by \begin{equation}\begin{split}
&  r^2\frac{\pa}{\pa a} \left\{\frac{Ta}{8\pi^{\frac{3}{2}}}\int_0^{\infty} t^{ \frac{1}{2}} \sum_{k=1}^{\infty} \sum_{l=0}^{\infty}\sum_{p=-\infty}^{\infty} \exp\left\{-\frac{1}{t}\left(  \omega_{N, l}^2 +\left[ 2\pi p T\right]^2\right)-t k^2a^2 \right\}  dt\right\}\\
\end{split}
\end{equation}
behaves as
\begin{equation}\label{eq1_12_3}\begin{split}
 r^2R^n &\Biggl\{-\frac{(n+2)\Gamma\left(\frac{n+3}{2}\right)\zeta_R(n+3)}{(4\pi)^{\frac{n+3}{2}}}\frac{T}{a^{n+3}}+\frac{2^{\frac{1-n}{2}} T^{\frac{n+5}{2}}}{a^{\frac{n+3}{2}}}
\sum_{k=1}^{\infty}\sum_{p=1}^{\infty}\left(\frac{p}{k}\right)^{\frac{n+3}{2}}K_{\frac{n+3}{2}}\left(4\pi kp Ta\right)\\&-\frac{2^{\frac{5-n}{2}}\pi T^{\frac{n+7}{2}}}{a^{\frac{n+1}{2}}}\sum_{k=1}^{\infty}\sum_{p=1}^{\infty} \frac{p^{\frac{n+5}{2}}}{k^{\frac{n+1}{2}}} K_{\frac{n+5}{2}}\left(4\pi kp Ta\right)\Biggr\}
+O(r^2R^{n-1}).
\end{split}
\end{equation}Using \eqref{eq1_12_2}, we find that \eqref{eq1_12_3} can also be written as\begin{equation}\label{eq1_12_4}\begin{split}r^2R^n&\Biggl\{-\frac{(n+3)\Gamma\left(\frac{n+4}{2}\right)\zeta_R(n+4)}{(4\pi)^{\frac{n+4}{2}} a^{n+4}}-
\frac{\Gamma\left(\frac{n+4}{2}\right)\zeta_R(n+4)}{\pi^{\frac{n+4}{2}}} T^{n+4} \\&+\frac{\pi T^{\frac{n+1}{2}}}{2^{\frac{n+1}{2}}a^{\frac{n+7}{2}}}\sum_{k=1}^{\infty}\sum_{p=1}^{\infty} \frac{k^{\frac{n+5}{2}}}{p^{\frac{n+1}{2}}}K_{\frac{n+1}{2}}\left(\frac{\pi kp}{Ta}\right) \Biggr\}+O(r^2R^{n-1}).
 \end{split}\end{equation}

\vspace{1cm}\noindent \textbf{Acknowledgments}\;
This project is   funded by Ministry of Science, Technology and Innovation, Malaysia under e-Science fund 06-02-01-SF0080.

\end{document}